\documentclass[aps,prl,twocolumn,superscriptaddress,showpacs,showkeys]{revtex4-1}
\usepackage[]{graphicx}
\begin{document}

\title{Probing the $N=32$ shell closure below the magic proton number $Z=20$:\\ Mass measurements of the exotic isotopes $^{52,53}\textrm{K}$}

\author{M. Rosenbusch}
\affiliation{Institut f\"ur Physik, Ernst-Moritz-Arndt Universit\"at Greifswald, 17487 Greifswald, Germany}

\author{P. Ascher}
\affiliation{Max-Planck-Institut f\"ur Kernphysik, Saupfercheckweg 1, 69117 Heidelberg, Germany}

\author{D. Atanasov}
\affiliation{Max-Planck-Institut f\"ur Kernphysik, Saupfercheckweg 1, 69117 Heidelberg, Germany}

\author{C. Barbieri}
\affiliation{Department of Physics, University of Surrey, Guildford GU2 7XH, UK}

\author{D. Beck}
\affiliation{GSI Helmholtzzentrum f\"ur Schwerionenforschung GmbH, Planckstra\ss e 1, 64291 Darmstadt, Germany}

\author{K. Blaum}
\affiliation{Max-Planck-Institut f\"ur Kernphysik, Saupfercheckweg 1, 69117 Heidelberg, Germany}

\author{Ch. Borgmann}
\affiliation{Max-Planck-Institut f\"ur Kernphysik, Saupfercheckweg 1, 69117 Heidelberg, Germany}

\author{M. Breitenfeldt}
\affiliation{Instituut voor Kern- en Stralingsfysica, KU Leuven, Celestijnenlaan 200d, B-3001 Heverlee, Belgium}

\author{R.B. Cakirli}
\affiliation{Max-Planck-Institut f\"ur Kernphysik, Saupfercheckweg 1, 69117 Heidelberg, Germany}
\affiliation{Department of Physics, University of Istanbul, 34134 Istanbul, Turkey}

\author{A. Cipollone}
\affiliation{Department of Physics, University of Surrey, Guildford GU2 7XH, UK}

\author{S. George}
\affiliation{Max-Planck-Institut f\"ur Kernphysik, Saupfercheckweg 1, 69117 Heidelberg, Germany}

\author{F. Herfurth}
\affiliation{GSI Helmholtzzentrum f\"ur Schwerionenforschung GmbH, Planckstra\ss e 1, 64291 Darmstadt, Germany}

\author{M. Kowalska}
\affiliation{CERN, 1211 Geneva 23, Switzerland}

\author{S. Kreim}
\affiliation{Max-Planck-Institut f\"ur Kernphysik, Saupfercheckweg 1, 69117 Heidelberg, Germany}
\affiliation{CERN, 1211 Geneva 23, Switzerland}

\author{D. Lunney}
\affiliation{CSNSM-IN2P3-CNRS, Universit\'e Paris-Sud, 91405 Orsay, France}

\author{V. Manea}
\affiliation{CSNSM-IN2P3-CNRS, Universit\'e Paris-Sud, 91405 Orsay, France}

\author{P. Navr\'atil}
\affiliation{TRIUMF, 4004 Westbrook Mall, Vancouver, BC, V6T 2A3, Canada}

\author{D. Neidherr}
\affiliation{GSI Helmholtzzentrum f\"ur Schwerionenforschung GmbH, Planckstra\ss e 1, 64291 Darmstadt, Germany}

\author{L. Schweikhard}
\affiliation{Institut f\"ur Physik, Ernst-Moritz-Arndt Universit\"at Greifswald, 17487 Greifswald, Germany}

\author{V. Som\`a}
\affiliation{CEA-Saclay, IRFU/Service de Physique Nucl\'eaire, 91191 Gif-sur-Yvette, France}
\affiliation{Institut f\"ur Kernphysik, Technische Universit\"at Darmstadt, 64289 Darmstadt, Germany}
\affiliation{ExtreMe Matter Institute EMMI, GSI Helmholtzzentrum f\"ur Schwerionenforschung GmbH, 64291 Darmstadt, Germany}

\author{J. Stanja}
\affiliation{Institut f\"ur Kern- und Teilchenphysik, Technische Universit\"at Dresden, 01069 Dresden, Germany}

\author{F. Wienholtz}
\affiliation{Institut f\"ur Physik, Ernst-Moritz-Arndt Universit\"at Greifswald, 17487 Greifswald, Germany}

\author{R. N. Wolf}

\affiliation{Institut f\"ur Physik, Ernst-Moritz-Arndt Universit\"at Greifswald, 17487 Greifswald, Germany}
\affiliation{Max-Planck-Institut f\"ur Kernphysik, Saupfercheckweg 1, 69117 Heidelberg, Germany}

\author{K. Zuber}
\affiliation{Institut f\"ur Kern- und Teilchenphysik, Technische Universit\"at Dresden, 01069 Dresden, Germany}


\begin{abstract}
The recently confirmed neutron-shell closure at $N=32$ has been investigated for the first time below the magic proton number $Z=20$ with mass measurements of the exotic isotopes $^{52,53}$K, the latter being the shortest-lived nuclide investigated at the online mass spectrometer ISOLTRAP. The resulting two-neutron separation energies reveal a $3\,\mathrm{MeV}$ shell gap at $N=32$, slightly lower than for $^{52}$Ca, highlighting the doubly-magic nature of this nuclide. Skyrme-Hartree-Fock-Boguliubov and \textit{ab initio} Gorkov-Green function calculations are challenged by the new measurements but reproduce qualitatively the observed shell effect.
\end{abstract}


\maketitle

\begin{figure*}
	\includegraphics[width=0.8\textwidth]{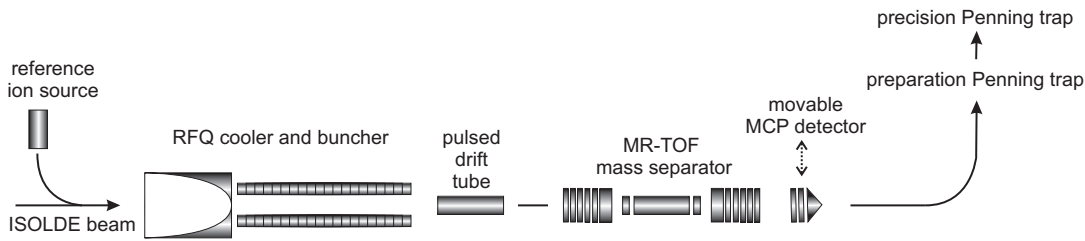}
	\caption{\label{setup} Sketch of the part of the ISOLTRAP setup used for the measurements reported here. For details, see text.}
\end{figure*}
Since the introduction of the shell concept in nuclear physics, the evolution of the shell structure far away from stability is one of the main research efforts. Similarly to atomic electrons, protons and neutrons can be regarded in a simple picture as occupying nuclear orbitals, of varying energy and angular momentum.  Nuclides show spherical nature and enhanced stability when these orbitals are filled leading to the so-called ``magic'' numbers of nucleons, however the standard shell picture has been shown to no longer be applicable to some exotic species. As magic numbers disappear going farther from stability, new shell closures arise in consequence, as in the case of the doubly magic oxygen isotope $^{24}$O \cite{Kanugo2009,Hoffman2009}. Such surprising isospin-dependent migrations of the nuclear quantum states point to the need to improve the understanding of the nuclear force \cite{Warner2004,Sorlin2008,Sorlin2013}.\\%
High excitation energies of the lowest $2^+$ excited state are a traditional clue for closed-shell nuclides, however only for even-even species. Two-neutron separation energies ($S_{2n}$) are another strong indication \--- even in the case of odd-$Z$ nuclides (such as potassium) and are obtained from mass measurements \cite{Blaum2013a,Schweikhard2006} as performed with mass spectrometers like the ISOLTRAP experiment at ISOLDE/CERN \cite{Mukherjee2008}.\\%
Beta-decay studies of $^{52}$Ca at ISOLDE published in 1985 \cite{Huck1985} revealed the energy of what was thought to be the first $2^+$ excited state (confirmed in 2006 \cite{Perrot2006}) and hinted at an enhanced $N=32$ shell effect. Recent mass measurements including $^{53,54}$Ca revealed a $4\,\mathrm{MeV}$ drop in $S_{2n}$ between $^{52}$Ca and $^{54}$Ca yielding a correspondingly large empirical shell gap \cite{Wienholtz2013}. An important test of the strength of the $N = 32$ shell closure comes from neighboring elements where proton correlations become active. For the isotopic chains of the heavier even-$Z$ elements titanium ($Z=22$), chromium ($Z=24$) and iron ($Z=26$), the $E(2^+)$ values have a local peak at $N=32$ although with an absolute magnitude lower than for $^{52}$Ca \cite{ENSDF14,Janssens2002}. Their two-neutron separation energies, however, as well as the ones of the odd-$Z$ neighbor elements, which have no $E(2^+)$ indicator, show no significant kink at $N=32$ \--- in sharp contrast to the closure of the $Z = 20$ proton core in calcium, where recently also for $^{54}$Ca a large $E(2^+)$ value has been revealed \cite{Steppenbeck2013}. Until now, the region below $Z=20$ was unexplored as these extremely exotic nuclides could not be produced in sufficient quantity.\\%
In this Letter, we present the first measurement of the masses of $^{52,53}$K, and thus the first investigation of the $N = 32$ shell closure towards the neutron dripline for $Z<20$. The experimental $S_{2n}$ values are compared to the predictions of self-consistent \textit{ab initio} Gorkov-Green function (GGF) theory. This recent development significantly extends the reach of calculations based on realistic interactions, including three-nucleon forces, in particular to open-shell (and odd-$Z$) nuclei. In contrast, we show that the predictions of mean-field calculations with phenomenological interactions are highly dependent on the parameterization of both the particle-particle and particle-hole parts of the energy functional. To this end, we performed in addition Hartree-Fock-Bogoliubov (HFB) calculations with the Skyrme interactions SLy4 and SLy5 \cite{Chabanat1998}, the latter containing a phenomenological tensor term.\\%

\begin{figure}
	\includegraphics[width=0.45\textwidth]{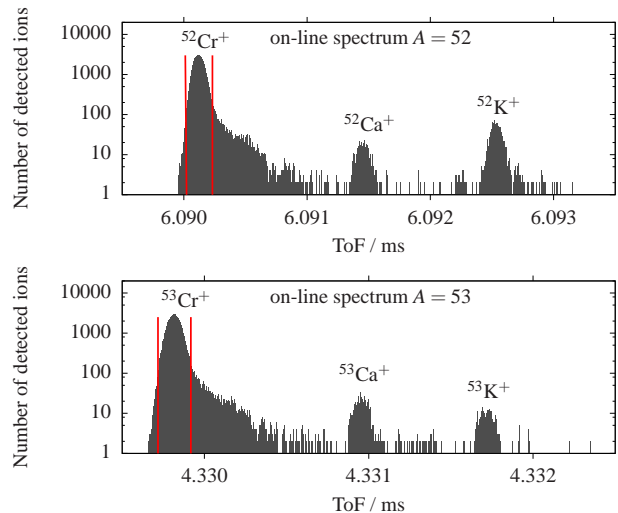}
	\caption{Time-of-flight spectra for nuclides at mass numbers $A=52$ (top) and $A=53$ (bottom). Vertical lines indicate the Cr$^+$ fit ranges.}
	\label{spectra}
\end{figure}

\begin{figure}
	\includegraphics[width=0.45\textwidth]{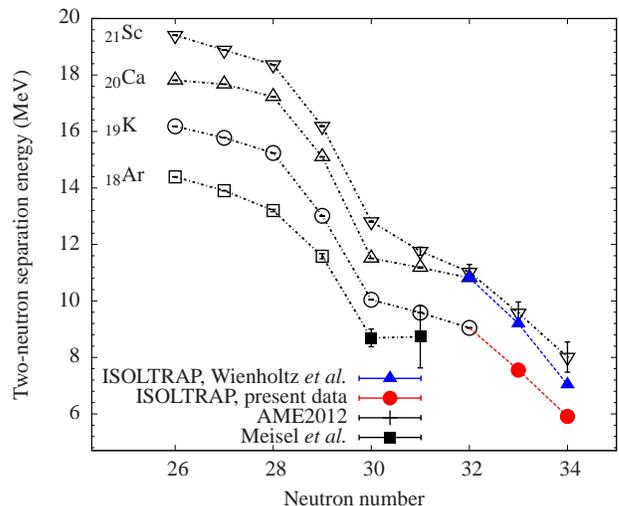}
	\caption{Two-neutron separation energies for the $Z=18-21$ isotopes from the atomic-mass evaluation 2012 (AME2012, open symbols) \cite{Wang2012}, recent measurements of $^{48,49}$Ar by Meisel \textit{et al.} \cite{Meisel2015}, and the ISOLTRAP data (from Ref. \cite{Wienholtz2013} and this work).}
	\label{S2n_1}
\end{figure}

\begin{figure}
	\includegraphics[width=0.45\textwidth]{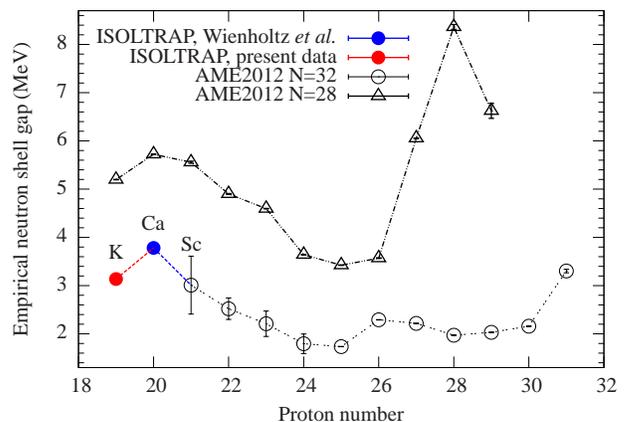}
	\caption{Empirical shell gaps for $N=28$ and $N=32$ from the atomic-mass evaluation 2012 (AME2012, open symbols) \cite{Wang2012}, and the ISOLTRAP data (from Ref. \cite{Wienholtz2013} and this work).}
	\label{shell-gap}
\end{figure}

\begin{figure}
\includegraphics[width=0.45\textwidth]{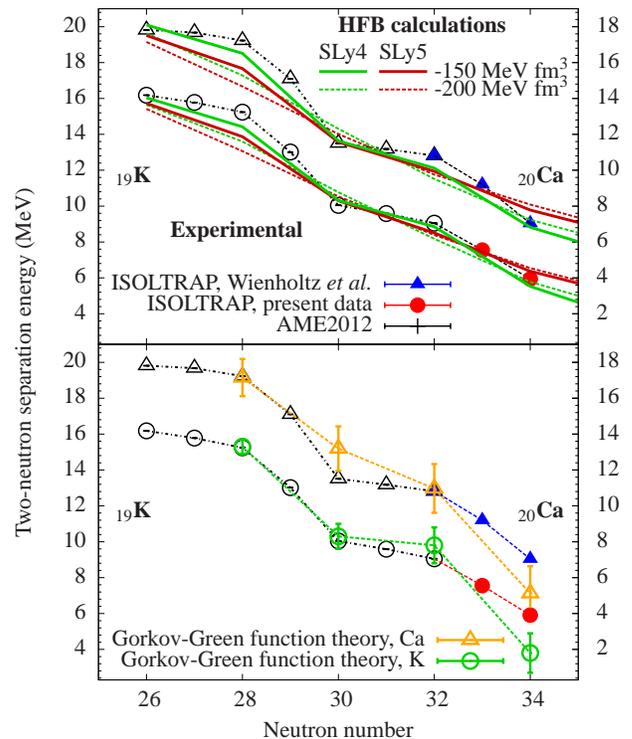}
\caption{Two-neutron separation energies for the isotopic chains of potassium (left axes) and calcium (right axes), note the different scales. Open symbols: data from \cite{Wang2012}, filled symbols: calcium data from \cite{Wienholtz2013} and new mass data from this work. Top: with $S_{2n}$ values from HFB calculations using the SLy4 (green lines) and the SLy5 (red lines) interaction, with volume-type delta pairing of strength $V_0=-150\,\mathrm{MeV\,fm}^3$ (solid lines) or $V_0=-200\,\mathrm{MeV\,fm}^3$ (dashed lines). Bottom: with $S_{2n}$ values obtained from \textit{ab initio} Gorkov-Green function theory (see text for details).}
\label{S2n_2}
\end{figure}

The exotic potassium isotopes were studied at ISOLDE/CERN with ISOLTRAP's multi-reflection time-of-flight mass separator (MR-ToF MS) \cite{Wolf2012a,Wolf2013a}. Figure~\ref{setup} shows the parts of the setup relevant for the present measurements: a radio-frequency quadrupole (RFQ) ion cooler and buncher \cite{Herfurth2001}, the MR-ToF MS, and a micro-channel plate (MCP) detector.\\%
The nuclides $^{52}$K and $^{53}$K were produced by impact of $1.4\textnormal{-}\mathrm{GeV}$ protons on a uranium carbide target. The atoms were surface ionized, along with stable chromium isobars. All ions were accelerated to $30\,\mathrm{keV}$, transported via the ISOLDE magnetic high-resolution mass separator (HRS) to the ISOLTRAP setup \cite{Mukherjee2008,Kreim2013a}, captured and cooled in the RFQ buncher, and ejected towards the MR-ToF MS. The energy of the ions was adapted by a pulsed drift tube prior to the injection into the MR-ToF MS, where in-trap-lift capturing was applied \cite{Wolf2012}. After several hundred reflections the ions were ejected towards the MCP detector.\\%
Figure~\ref{spectra} shows time-of-flight spectra of isobaric ions at mass numbers $A=52$ and $A=53$. The calcium peaks were identified by an element-selective laser-ionization scheme. For ToF signals at longer flight times there are no known isobaric atomic ions except $^{52,53}$K. The expected flight times of possible molecular ions produced at ISOLDE, such as oxides, dioxides, sulfides, hydrides, and of doubly charged ions have been compared with the observed peaks. The only species that would come close are $^{36,37}$Si($^{16}$O)$^+$, that contain short-lived silicon isotopes. However, due to the slow chemical release from the target, the probability for short-lived silicon beam components is very low. More importantly, it can be excluded as only in the case of $A=52$ the flight time of $^{36}$Si($^{16}$O)$^+$ would fit the measured peak; at $A=53$, the measured mass deviates by $1000(130)\,\mathrm{keV}$ from the value of $^{37}$Si($^{16}$O)$^+$ \cite{Wang2012}, which corresponds to a time-of-flight difference in the order of $40\,\mathrm{ns}$.\\%
In four hours, 12000 counts were collected for $^{52}$K, and in 12 hours, 2300 counts for $^{53}$K. Note that with a half-life of just $30\,\mathrm{ms}$, $^{52}$K is the shortest-lived nuclide ever investigated with ISOLTRAP. For mass calibration, stable $^{39}$K from the ISOLTRAP reference ion source, as well as $^{52,53}$Cr ions from ISOLDE were used. With the reference-ion masses $m_{1,2}$ and their corresponding flight times $t_{1,2}$, the ion mass of interest $m$ is determined from its time of flight $t$ by \cite{Wienholtz2013}:%
\begin{equation}
m^{1/2} = C_{ToF}(m_1^{1/2}-m_2^{1/2})+(m_1^{1/2}+m_2^{1/2})/2,%
\end{equation}
where the experimental values are described by the single variable \mbox{$C_{ToF} = (2t-t_1-t_2)/\left[2(t_1-t_2)\right]$}. Due to the deviation from Gaussian shape of the intense chromium ion peaks as shown in Fig.~\ref{spectra}, the mean flight times have been determined from a limited region around the peak maximum (as indicated by vertical lines) that cover about $95\,\mathrm{\%}$ of the measured events. Uncertainties that arise from different choices of this region have been investigated and taken into account.\\
	\begin{table}
	\caption{Half-lives $T_{1/2}$ \cite{Wang2012}, reference nuclides (Ref. nucl.), experimental ToF variable $C_{ToF}$, and resulting mass-excess ($ME$) values of the investigated isotopes $^{52,53}$K. The mass values of the reference nuclides are taken from \cite{Wang2012}.\\}
		\begin{tabular}{ccccc}
			Isotope&$T_{1/2}$&Ref. nucl.&$C_{ToF}$&$ME$ (keV)\\\hline
			$^{52}$K&$110(4)$\,ms&$^{39}$K,$^{52}$Cr&$0.50295425(256)$&$-17138(33)$\\
			$^{53}$K&$30(5)$\,ms&$^{39}$K,$^{53}$Cr&$0.50306656(799)$&$-12298(112)$
		\end{tabular}{}
	\label{ME-table}
	\end{table}
The resulting mass excesses $ME=M-Au$ are listed in Tab. \ref{ME-table}, where $M$ is the atomic mass, $A$ is the mass number, and $u$ is the atomic mass unit. We note that, in principle, $^{52}$Ca, with its mass well known from recent Penning-trap measurements \cite{Wienholtz2013,Gallant2012}, would be the second mass reference of choice (in addition to $^{52}$Cr) for measuring $^{52}$K rather than $^{39}$K from the reference ion source as $^{52}$Ca appears in the same spectrum with $^{52}$K and $^{52}$Cr. In general, such a procedure (used here for the first time) is preferred because the MR-ToF data of all three isobaric ions are taken simultaneously, excluding drifts between the measurements. The mass values from both calibrations are in agreement. However, the isobaric calibration has a larger uncertainty due to significantly lower statistics of the $^{52}$Ca as compared to the $^{39}$K signal from the reference ion source.\\%
In Figure~\ref{S2n_1}, the two-neutron separation energies \mbox{$S_{2n} = B(Z,N) - B(Z,N-2)$}, where $B(N,Z)$ is the binding energy of a nuclide with $Z$ protons and $N$ neutrons, of the isotopic chains from argon to scandium are presented, including the present data. The new $S_{2n}$ values show a significant drop from $^{51}$K \cite{Gallant2012} to $^{53}$K: about $3\,\mathrm{MeV}$, compared to only about $1\,\mathrm{MeV}$ from $^{49}$K to $^{51}$K, clearly illustrating the $N=32$ shell effect for potassium. The large uncertainty of the $S_{2n}$ value of $^{53}$Sc ($271\,\mathrm{keV}$) \cite{Wang2012,Crawford2010} at $N=32$ leaves room for some ambiguity as to whether the $S_{2n}$ values of calcium and scandium cross. Thus, new precision-mass measurements of the corresponding Sc isotopes would be desirable.\\%
In Fig.~\ref{shell-gap} the empirical two-neutron shell gap \mbox{$S_{2n}(N,Z)-S_{2n}(N+2,Z)$} is plotted as a function of the proton number for $N=28$ and $N=32$, including the new point below the magic proton number $Z=20$. A local maximum of the $N=32$ shell gap has thus been revealed for the proton-magic $Z=20$ core. In addition, the plot shows a similar behavior for both neutron numbers going towards the dripline with lower $Z$, where the shell gap is rising up to a maximum at $Z=20$ and falling by about $1\,\mathrm{MeV}$ at $Z=19$.\\ 
As shown already in \cite{Wienholtz2013} for the calcium isotopes, Density-Functional-Theory (DFT) calculations reproduce the average $S_{2n}$ trend across $N = 28$ and $N = 32$, but do not predict the sharp drops associated with the two shell closures. We extend this study to $Z<20$ and perform Skyrme-Hartree-Fock-Bogoliubov (Skyrme-HFB) \cite{Ring2000a} calculations using the HFODD code \cite{Schunck2012} and interactions of the Skyrme-Lyon family \cite{Chabanat1998}, in the spherical approximation.\\%
The calculations use the SLy4 or SLy5 interactions for the particle-hole part and a volume delta-pairing interaction for the particle-particle part of the energy functional, the latter being defined exclusively by a strength parameter $V_0$. With respect to the SLy4 interaction, SLy5 contains in addition a tensor term.\\%
The top panel of Figure~\ref{S2n_2} shows the experimental and computed HFB $S_{2n}$ values for the potassium and calcium isotopic chains. The theoretical $S_{2n}$ are computed for nuclei of even neutron number. Self-consistent quasi-particle blocking of the odd protons is performed for the potassium isotopes, using the procedure described in \cite{Dobaczewski2009}.\\%
A strength of the pairing interaction of $-200\,\mathrm{MeV\,fm}^3$ reproduces the very smooth $S_{2n}$ trend observed in \cite{Wienholtz2013}. It describes correctly the experimental values on average but underestimates the drop at the crossing of the magic neutron numbers. A reduction of the strength of the pairing interaction (solid lines) leads to a significant improvement of the description of the experimental $S_{2n}$ trend. The addition of the tensor term with the SLy5 interaction leads to a change in the wrong direction. However, a recent work \cite{Grasso2014} has shown that the effect of the tensor term in mean-field calculations strongly depends on the way it is constrained to experimental data.

In addition to the empirical HFB approach, it is now possible to perform calculations up to the medium mass region using \textit{ab initio} methods (see, e.g., Refs. \cite{Hergert2013,Cipollone2013,Soma2014b,Hagen2008,Babieri2009,Hagen2012,Hergert2014,Binder2014}). Thus, new mass calculations have been performed in the \textit{ab initio} Gorkov-Green function (GGF) framework \cite{Soma2011,Soma2014a,Soma2014b} that allows for the study of open-shell nuclei. This method is particularly suited for the present purpose due to the ease of calculating odd-even systems, which also makes it a unique tool to investigate neighboring isotopic chains.\\%
In our calculations the only input are two- and three-body interactions fitted to properties of systems with $A=2,3$ and $4$, without any further adjustments of the parameters. GGF calculations have recently addressed the region around $Z=20$ \cite{Soma2014b} and are extended here for the first time beyond N = 32 for potassium.\\%
The present calculations made use of two- and three-nucleon forces derived within chiral effective field theory at next-to-next- and next-to-next-to-next-to-leading order (N$^2$LO and N$^3$LO), respectively \cite{Entem2003,Navratil2007}, extended to the low-momentum scale $\lambda=2.0\,\mathrm{fm}^{-1}$ by means of free-space similarity renormalization-group techniques. The many-body treatment is set by a second-order truncation in the GGF self-energy expansion \cite{Soma2011}. 
Model spaces up to 14 harmonic oscillator shells were employed and three-body interactions were restricted to basis states with $E_{\text{3max}} \leq 16$. Infrared extrapolations of the calculated ground state energies were subsequently performed following Ref. \cite{More2013}. We note that, in the present case, this procedure is formally defective due to the different truncations of one- and three-body model spaces. Nevertheless, we find that the trend expected from Ref. \cite{More2013} is qualitatively reproduced, although with larger extrapolation uncertainties. This is in agreement with other calculations \cite{Hergert2014}. As an example, we obtain binding energies of $439.52(0.71)\,\mathrm{MeV}$ for $^{51}$K and $443.31(0.85)\,\mathrm{MeV}$ for $^{53}$K. This overbinding of about $0.7\,\mathrm{MeV/A}$ is a general feature of currently available chiral interactions and it is a constant effect throughout the whole isotopic chain that cancels in separation energies \cite{Soma2014b,Hergert2014,Binder2014}.\\%
GGF results for $S_{2n}$ of $^{47,49,51,53}$K and $^{48,50,52,54}$Ca are shown in the bottom panel of Fig.~\ref{S2n_2} and are all resulting from the infrared extrapolation. Different sources of uncertainty affect the present theoretical results (see Refs. \cite{Soma2014a,Soma2014b} for a detailed discussion). In particular, this method breaks particle-number symmetry (like HFB theory) and generates the correct expectation values for the proton and neutron numbers only on average, with a finite variance. However, the associated errors are expected to cancel with good accuracy for energy differences (such as $S_{2n}$). The errors indicated are uniquely those originating from the extrapolation fit and range between $0.4$ to $1.5\,\mathrm{MeV}$ with increasing mass number. In general, GGF calculations are in fair agreement with measured $S_{2n}$, with the mismatch at $^{53}$K being on the order of the truncation error. The significant drop from $^{51}$K to $^{53}$K is qualitatively reproduced but overestimated by theory, which also leads to an overestimation of the empirical shell-gap for potassium. In contrast to the $N=28$ gap, which is quantitatively better reproduced, the overestimation of the $N=32$ gap emerges as a common feature of \textit{ab initio} calculations in this mass region (see also \cite{Hergert2014}). The present measurements therefore constitute an important test for the on-going developments of chiral interactions and \textit{ab initio} methods.

In summary, direct mass measurements performed with ISOLTRAP's multi-reflection time-of-flight mass separator show that the neutron-shell closure at $N=32$ extends below the closed proton shell $Z=20$. The empirical gap is about $1\,\mathrm{MeV}$ weaker than for calcium indicating the stabilizing effect of the $Z=20$ proton core and highlighting the doubly-magic nature of $^{52}$Ca. Mean-field calculations for K and Ca show that interactions with effective parameters, although leading to more easily accessible calculations, depend on the way the parameters are fitted and are thus not always precisely enough constrained to yield robust predictions of shell effects. In contrast, Gorkov-Green function calculations, extended to $N>32$ along the potassium chain for the first time, offer predictions that are essentially parameter free and without any fitting to this mass region. By providing novel details on shell evolution, the measurements represent an important test and are instrumental for validating \textit{ab initio} theories and extending them to heavier and more exotic nuclei.%
\begin{acknowledgments}%
We thank T. Duguet and A. Gottberg for stimulating discussions. This work was supported by the German Federal Ministry for Education and Research (BMBF) (contract no. 05P12HGCI1, 05P12HGFNE and 05P09ODCIA), the Max-Planck Society, the European Union seventh framework through ENSAR (contract no. 262010), the French IN2P3, the ISOLDE Collaboration, the DFG through Grant No. SFB 634, by the Helmholtz Alliance Program, Contract No. HA216/EMMI, by the United Kingdom Science and Technology Facilities Council (STFC) under Grants ST/L005743/1 and ST/L005816/1 and by the Natural Sciences and Engineering Research Council of Canada (NSERC), Grant No. 401945-2011.\\%
GGF calculations were performed using HPC resources from GENCI-CCRT (Grants No. 2014-050707 and 2015-057392) and the DiRAC Data Analytic system at the University of Cambridge (BIS National E-infrastructure capital grant No. ST/K001590/1 and STFC grants No. ST/H008861/1, ST/H00887X/1, and ST/K00333X/1). SK acknowledges support from the Robert-Bosch Foundation.
\end{acknowledgments}

\end{document}